\documentclass[journal]{IEEEtran}

\ifCLASSINFOpdf
\usepackage[pdftex]{graphicx}
\DeclareGraphicsExtensions{.pdf,.jpeg,.png}
\else
\usepackage[dvips]{graphicx}
\fi
\usepackage{subfigure}
\usepackage{epstopdf}
\usepackage[cmex10]{amsmath}
\usepackage{amssymb}
\usepackage{wasysym}
\usepackage{relsize}
\usepackage{algorithm}
\usepackage{algorithmic}
\usepackage{array}
\usepackage{cite}
\usepackage{color}
\usepackage{xcolor}
\usepackage{url}
\usepackage{bm}
\usepackage{enumerate}
\usepackage{epsfig,latexsym}%amsmath
\usepackage{cite}

\usepackage{multirow}
\usepackage{threeparttable}
\usepackage{makecell}
\usepackage{setspace}

\usepackage{lipsum} % For dummy text, you can remove this in your actual document

\makeatother

\begin{document}
\title{Near-Field Communications for Extremely Large-Scale MIMO: A Beamspace Perspective}

\author{Kangjian~Chen,~\IEEEmembership{Student~Member,~IEEE}, Chenhao~Qi,~\IEEEmembership{Senior~Member,~IEEE}, \\ Jingjia~Huang,~\IEEEmembership{Student~Member,~IEEE}, Octavia A. Dobre,~\IEEEmembership{Fellow,~IEEE} and Geoffrey Ye Li,~\IEEEmembership{Fellow,~IEEE}
\thanks{Kangjian~Chen, Chenhao~Qi and Jingjia Huang are with the School of Information Science and Engineering, Southeast University, Nanjing 210096, China (e-mail: \{kjchen, qch, jiah\}@seu.edu.cn). Octavia A. Dobre is with the Faculty of Engineering and Applied Science, Memorial University, St. John’s, NL A1C 5S7, Canada (e-mail: odobre@mun.ca). Geoffrey Ye Li is with the Department of Electrical and Electronic Engineering, Imperial College London, SW7 2AZ London, U.K. (e-mail:	geoffrey.li@imperial.ac.uk). (\textit{Corresponding author: Chenhao~Qi})
}
}

\markboth{Accepted by IEEE Communications Magazine}
{}
\maketitle
\begin{abstract}
Extremely large-scale multiple-input multiple-output (XL-MIMO) is regarded as one of the key techniques to enhance the performance of future wireless communications. Different from regular MIMO, the XL-MIMO shifts part of the communication region from the far field to the near field, where the spherical-wave channel model cannot be accurately approximated by the commonly-adopted planar-wave channel model. As a result, the well-explored far-field beamspace is unsuitable for near-field communications, thereby requiring the exploration of specialized near-field beamspace. In this article, we investigate the near-field communications for XL-MIMO from the perspective of beamspace.  Given the spherical wavefront characteristics of the near-field channels, we first map the antenna space to the near-field beamspace with the fractional Fourier transform. Then, we divide the near-field beamspace into three parts, including high mainlobe, low mainlobe, and sidelobe, and provide a comprehensive analysis of these components. Based on the analysis, we demonstrate the advantages of the near-field beamspace over the existing methods. Finally, we point out several applications of the near-field beamspace and  highlight some potential directions for future study in the near-field beamspace.
\end{abstract}

\renewcommand{\abstractname}{Research Abstract}
\begin{abstract}
	The authors investigate the near-field communications for extremely large-scale multiple-input multiple-output from a beamspace perspective.
\end{abstract}

%\newcommand{\researchabstractname}{Research Abstract}
%\newenvironment{researchabstract}{%
%	\small
%	\begin{center}%
%		{\bfseries \researchabstractname\par}%
%	\end{center}%
%	\quotation
%}{\endquotation}
%
%\begin{researchabstract}
%	The authors investigate the near-field communications for extremely large-scale multiple-input multiple-output from a beamspace perspective.
%\end{researchabstract}

%\begin{IEEEkeywords}
%Extremely large-scale multiple-input multiple-output (XL-MIMO), hardware implementation, near-field beam pattern, near-field communications
%\end{IEEEkeywords}

\begin{figure*}[htbp]
	\centering
	\includegraphics[width=170mm]{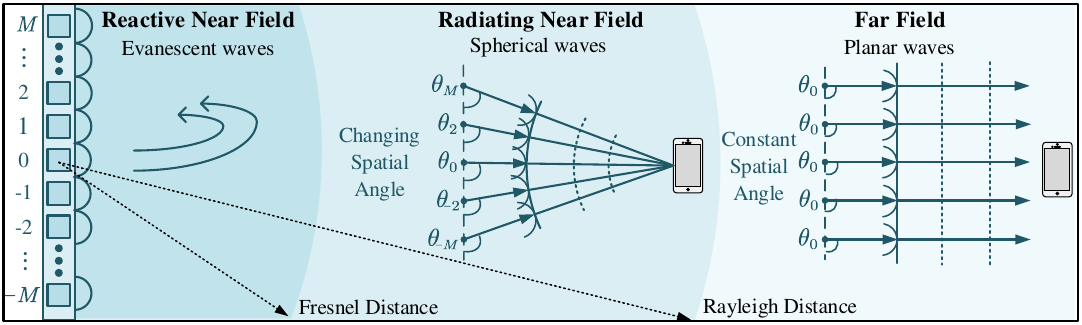}
	\caption{Illustration of characteristics of the near and far fields.}
	\label{NearFieldFarField}
	\vspace{-0.5cm}
\end{figure*}

\section{Introduction}
Future wireless communications are expected to support many applications, such as smart cities and vehicles to everything, that often demand ultrahigh-speed transmission or ultra-dense connections. To enable these applications, extremely large-scale multiple-input multiple-output (XL-MIMO) is developed as a candidate technique~\cite{TcomLYH,TWC24YZQ,CL23JGL}. In XL-MIMO, a base station (BS) typically adopts an antenna array with far more antennas than massive MIMO, e.g., 1024 antennas vs. 64 antennas. As a result, the BS can not only  provide high beamforming gains to improve the spectral efficiency but also utilize the increased spatial degree of freedom to enhance the system connectivity~\cite{TcomLYH} .

On the other side, the considerable increase in array aperture has differentiated channel models of XL-MIMO from those of the massive MIMO. According to the array aperture and the communication distance, the electromagnetic wave radiation field is categorized into the far field, radiating near field, and reactive near field, as depicted in Fig.~\ref{NearFieldFarField}. In massive MIMO, the limited array aperture results in a far-field-dominated BS coverage, characterized by the planar-wave channel model. Conversely, due to XL-MIMO’s significant array expansion, the near field, encompassing both radiating and reactive components, constitutes a substantial part of the BS coverage, where the spherical-wave model is adopted for accurate channel characterization~\cite{TWC24YZQ}. The changes in channel models have fundamentally differentiated near-field communications from the existing far-field ones, which necessitates the development of dedicated signal processing techniques for near-field communications.

%By harnessing the characteristics of the spherical wavefront in the near field, the BS can concentrate beam energy on specific locations instead of directions, which is referred to as the beam focusing~\cite{TWC22ZHY}. 

In addition to the challenges posed by the difference in channel models, the XL-MIMO also suffers from the high system processing complexity due to the deployment of a large number of antennas. An effective way to reduce this complexity is the beamspace processing~\cite{BSMIMO}. By exploiting the channel sparsity and transforming the received signals from the antenna space to the beamspace, the signal dimensions are condensed, and consequently the processing complexity is reduced. Besides complexity reduction, beamspace processing has various other advantages, including alignment with hybrid precoding structures, improved signal-to-noise ratio, enhanced system robustness, spatial interference mitigation, and signal feature extraction. In the far field, the antenna space and the beamspace can be conveniently connected by the Fourier transform. Benefiting from  the classic Fourier transform theory, the far-field beamspace processing methods have been widely applied in numerous aspects, such as beam training, beam tracking, and channel estimation.  Given the enlarging number of antennas of XL-MIMO, there arises increasing demand for the development of beamspace processing. However, the fundamental difference between far-field  and near-field channel models renders the well-explored far-field beamspace unsuitable for the near field. Consequently, a crucial need arises for the exploration of beamspace specifically designed for near-field communications.

%Since the XL-MIMO is likely to operate in the coexistence of the near and far field, the beamspace processing for XL-MIMO should adapt to the hybrid field, including the far and near field.

%To this end, we unveil the beamspace characteristics of near-field channels by analyzing the beam patterns of near-field steering vectors in different domains in Fig.~\ref{NearfieldBeamFocusing}. 
%
%
%
%Inspired by the distinct channel models and diverse application potentials of the near field, in this article, 

%Most of the existing works perform the

The most widely adopted near-field beamspace processing method is the near-field beam focusing~\cite{TWC22ZHY}. This approach utilizes the spherical wavefront properties of the near field to concentrate beam energy at specific locations. However, the near-field beam focusing only concentrates on the region around the focused location and neglects other  extensive regions in the near field.  To date, there has been no comprehensive investigation of near-field communications from a beamspace perspective. To fill this gap, this article presents an in-depth discussion on near-field beamspace for XL-MIMO systems. Different from the far field that connects the antenna space and the beamspace by the Fourier transform, we first  map the antenna space to the near-field beamspace with the fractional Fourier transform (FrFT) according to the spherical wavefront characteristics of the near-field channels. We divide the near-field beamspace into three parts, including high mainlobe, low mainlobe, and sidelobe. Then, we provide a  detailed analysis of the high and low mainlobes in the near-field beamspace. Based on the analysis, we demonstrate the advantages of the near-field beamspace over the existing methods. Finally, we point out several applications of the near-field beamspace and  highlight some potential directions for future study in the near-field beamspace.

\section{Near-Field Beamspace}\label{NFChanenl}
In this section, we explore the near-field beamspace for XL-MIMO. We commence by discussing far-field and near-field channel models and introducing their  corresponding beamforming approaches. According to the spherical wavefront characteristics of the near-field channels,  we map the antenna space to the near-field beamspace with the FrFT. Then, we conduct a comprehensive investigation on the near-field beamspace.

%
%\subsection{Planar-  v.s. Spherical-Wave Models}
%Electromagnetic waves have a spherical wavefront when propagating through the free space. As a result, the phase differences and amplitude variations among antennas are modeled based on the distances between the transmit and receive antennas, leading to the ground-truth spherical-wave model. In this context, the channel coefficients exhibit a high nonlinearity with respect to antenna positions, which greatly increases the complexity of system implementation. For conventional wireless systems, due to the small array aperture, users are mainly located in the far field, where the spherical-wave model can be simplified as the planar-wave model, and phase differences among antennas are only related to the channel angle. The concise planar-wave model greatly simplifies the spherical-wave model and has facilitated extensive studies on conventional wireless systems. However, for the emerging XL-MIMO, a significant proportion of users are located in the near field due to the expansion in array aperture. If we continue to adopt the planar-wave model in the near field, a large performance loss will be incurred. Therefore, in the near field, we must stick to the original spherical-wave model to characterize the channel. In this condition,  the phase differences among antennas are related to both the channel angle and distance.

\begin{figure*}[!t]
	\centering
	\includegraphics[width=140mm]{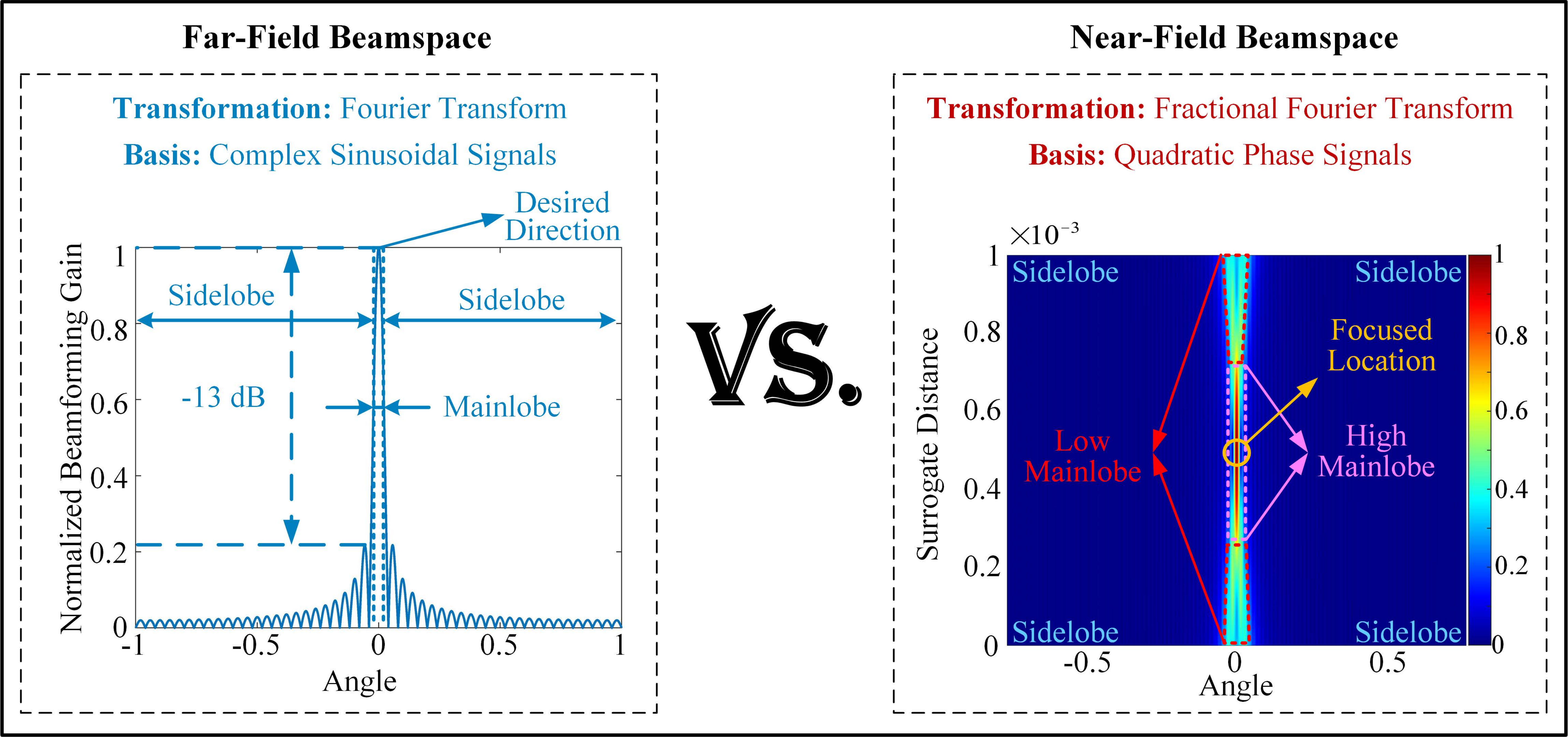}
	\caption{Comparisons of far-field and near-field beamspaces.}
	\label{BeamSpace}
	\vspace{-0.4cm}
\end{figure*}
% Conversely, if we adopt the ground-truth spherical-wave model, the complexity of system implementation would be unaffordable. Therefore, it is important to consider both the performance and complexity of NFC when dealing with near-field channels.

\subsection{Planar-  vs. Spherical-Wave Models}
Electromagnetic waves have a spherical wavefront when propagating through the free space. As a result, the phase differences and amplitude variations among antennas are modeled based on the distances between the transmit and receive antennas, leading to the ground-truth spherical-wave model. In this context, the channel coefficients exhibit a high nonlinearity with respect to antenna positions, which greatly increases the complexity of system implementation. For conventional wireless systems, due to the small array aperture, users are mainly located in the far field, where the spherical-wave model can be simplified as the planar-wave model, and phase differences among antennas are only related to the channel angle. The concise planar-wave model greatly simplifies the spherical-wave model and has facilitated extensive studies on conventional wireless systems. However, for the emerging XL-MIMO, a significant proportion of users are located in the near field due to the expansion in array aperture. If we continue to adopt the planar-wave model in the near field, a large performance loss will be incurred. Therefore, in the near field, we must stick to the original spherical-wave model to characterize the channel, where the phase differences among antennas are related to both the channel angle and distance.

\subsection{Far-Field Beam Steering vs. Near-Field Beam Focusing}
The different channel models in the far and near fields lead to different beamforming approaches. In general, based on the phase differences among antennas, beamforming can make full use of the power of the antenna array to combat the  propagation attenuation.  In the far field, where phase differences are primarily associated with channel angles, the beamforming directs energy to specific directions, leading to far-field beam steering. In contrast, the near field, characterized by a spherical wavefront, introduces phase differences related to both channel angle and distance. By carefully adjusting the beamforming weights to compensate the phase differences, the near-field beamforming can focus the energy on specific locations, leading to the near-field beam focusing~\cite{TWC22ZHY}. The distinct beamforming approach  in the near field results in some unique applications. For example, in the near-field multiuser communications, the BS can exploit the near-field beam focusing to serve users at the same angle but different distances. In addition, since electromagnetic waves exhibit a spherical wavefront in the near field,  the near-field steering vector depends on both the angle and distance, which enables near-field sensing and localization.

\subsection{Far- vs. Near-Field Beamspaces}
The deployment of a large number of antennas necessitates the beamspace processing to reduce the complexity or improve the performance. However, the far-field beam steering only concentrates on the angular interval around the desired direction and neglects the other directions. Inspired by the similarity between far-field channel steering vectors and complex sinusoidal signals, the Fourier transform has been introduced to map the antenna space to the far-field beamspace~\cite{BSMIMO}. As illustrated in  the left  of  Fig.~\ref{BeamSpace}, besides the desired direction, the far-field beamspace, which can be divided into the mainlobe and the sidelobe, reveals the overall characteristics of the far-field channel steering vector across all the directions. For example, the mainlobe exhibits an energy concentration within a specific angular interval, where the beamwidth  depends on the number of antennas. Furthermore, the sidelobe demonstrates significantly lower beamforming gain compared to the mainlobe, rendering it negligible in certain scenarios. This common viewpoint leads to the extensive adoption of far-field beamspace processing, promoting the advancement of massive MIMO.

Similarly,  the near-field beam focusing only concentrates on  the region around  the focused location and neglects the others.  Then, a problem arises naturally: Can we develop the corresponding  near-field beamspace beyond the near-field beam focusing?

\begin{figure*}[htbp]
	\centering
	\includegraphics[width=180mm]{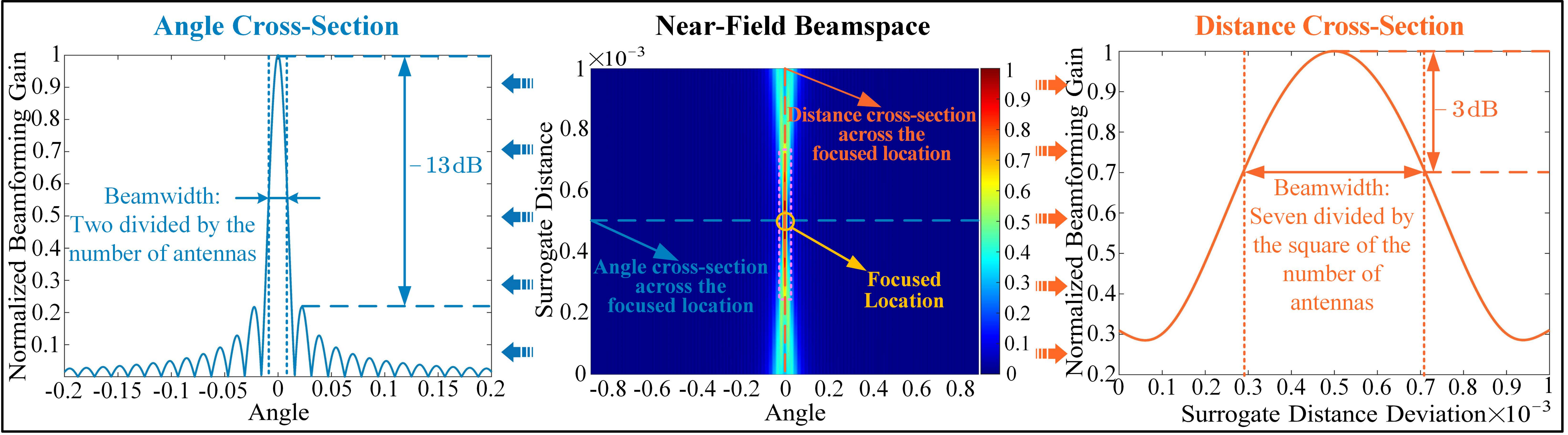}
	\caption{Characteristics of the high mainlobe in the near-field beamspace.}
	\label{NearfieldBeamFocusing}
			\vspace{-0.4cm}
\end{figure*}

\begin{figure}[!t]
	\centering
	\includegraphics[width=75mm]{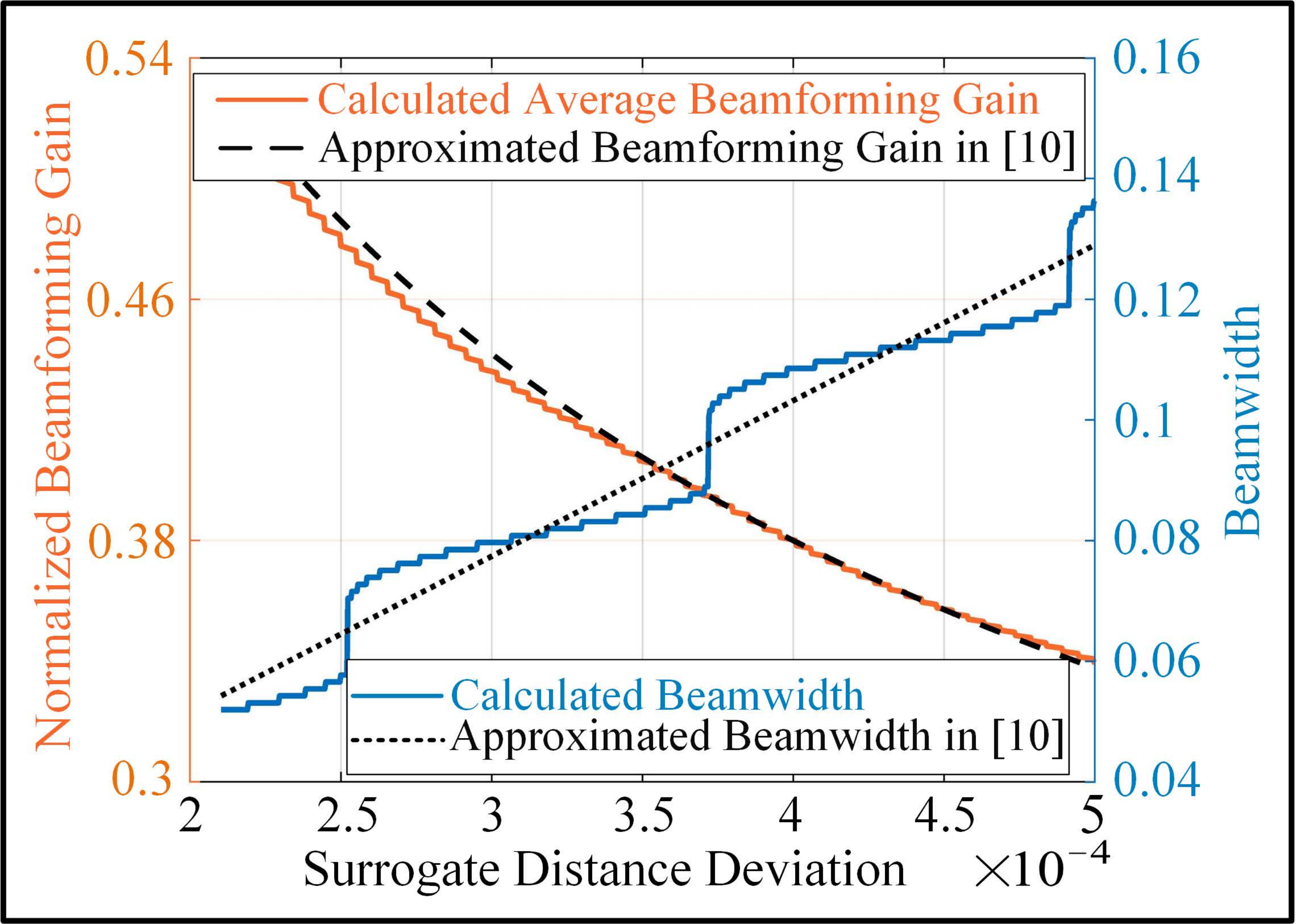}
	\caption{Spatial region and intensity distribution of the low mainlobe in the near-field beamspace.}
	\label{LowMainlobe}
			\vspace{-0.5cm}
\end{figure}

As shown in Fig.~\ref{NearFieldFarField}, in the far field, all antennas share the same spatial angle due to the planar wavefront characteristics.  However, in the near field, spherical wavefront characteristics  cause the spatial angles to vary across different antenna indices. As a result, the near-field channel steering vectors are more similar to the frequency-modulated signals than to the complex sinusoidal signals with constant frequencies. Specifically, by employing the Fresnel approximation~\cite{Tcom22CMH}, the phases of antennas are quadratic functions of the antenna indices, which are similar to the linear frequency modulation (LFM) signals or chirp signals~\cite{TWC23SX}. Since the Fourier transform typically  captures the overall frequency characteristics of a signal, it cannot characterize spherical wavefront characteristics of near-field channels, and therefore is not suitable for the near-field beamspace analysis. A common way to deal with the frequency-modulated signals in the existing literature is the FrFT~\cite{TSP94ALB}.  Different from the Fourier transform that uses the complex sinusoidal signals as the basis, the FrFT uses the quadratic phase signals as the basis, which matches perfectly with near-field channel steering vectors based on the Fresnel approximation. In light of this, we propose to map the antenna space to the near-field beamspace with the FrFT.

As illustrated in the right  of  Fig.~\ref{BeamSpace}, by performing the FrFT, we transform  the near-field channel steering vector from the antenna space to the near-field beamspace.  Before delving into the characteristics of the near-field beamspace,  we provide some clarifications.  The Fourier transform basis is the complex sinusoidal signal and only has one parameter. As a result, the far-field beamspace is two-dimensional with the angle and the beamforming gain being the x-axis and y-axis, respectively. However, the FrFT basis is the quadratic phase signal and has two parameters, which leads to the three-dimensional near-field beamspace. In addition, the linear and quadratic coefficients of the FrFT basis correspond to the angle and distance dimensions of the near-field channel steering vectors based on the Fresnel approximation, respectively. Consequently, we designate the coefficient of the quadratic term as the ``surrogate distance'' and the coefficient of the linear term as the ``angle''. As illustrated in  the right  of  Fig.~\ref{BeamSpace}, the x-axis and y-axis of the near-field beamspace are the angle and the surrogate distance, respectively. Therefore, every point in the near-field beamspace corresponds to a specific  location in the near field. Since complex sinusoidal signals exhibit linear phases and are special cases of quadratic phase signals with zero quadratic terms, the Fourier transform is a particular case of the FrFT. Consequently, the far-field beamspace is a subset of the near-field beamspace when the surrogate distance is zero. Therefore, the near-field beamspace can also be utilized to analyze far-field channels. When the basis of the FrFT aligns with the near-field steering vector, the energy can be concentrated on a specific location in the near field, i.e., the focused location in  the right  of  Fig.~\ref{BeamSpace}, which is named as the near-field beam focusing~\cite{TcomLYH,TWC22ZHY}. Besides the focused location, the near-field beamspace also demonstrates other characteristics. Notably, according to the beamforming gains, we divide the beamspace of a near-field channel steering vector into three parts: high mainlobe, low mainlobe, and sidelobe. It is evident that the beamforming gain of the sidelobe is significantly smaller than that of the mainlobe, encompassing both the high and low mainlobes.  Therefore, we focus our analysis on the mainlobes and omit the sidelobe.

%In addition, comparing the the basis for FrFT and that of the Fourier transform, we can find that the Fourier transform is a special FrFT with the surrogate distance equals zeros. Therefore, the far-field beamspace can be seen as a cross-section of the near-field beamspace.

\subsection{Detailed Analysis of the High and Low Mainlobes}\label{NFBA}
In addition to the focused location of the near-field beam focusing~\cite{TWC22ZHY}, analysis on the characteristics of both high and low mainlobes is crucial and will be detailed as follows.

\subsubsection{\textbf{High Mainlobe}} In Fig.~\ref{NearfieldBeamFocusing}, we analyze the characteristics of the high mainlobe in the near-field beamspace. In the middle  of  Fig.~\ref{NearfieldBeamFocusing}, we begin by duplicating  the near-field beamspace from  the right  of  Fig.~\ref{BeamSpace} and then mark the focused location, the angle cross-section across the focused location, and the distance cross-section across the focused location. 

\textit{Angle Cross-Section:} In the left of Fig.~\ref{NearfieldBeamFocusing}, we illustrate the angle cross-section of the near-field beamspace across the focused location.  Since the angle cross-section passes through the focused location, the surrogate distance of the FrFT basis aligns with the quadratic coefficient of the channel steering vector. In this context, the FrFT of the channel steering vector would become the Fourier transform of a far-field channel steering vector. As a result, the properties of the angle cross-section of the near-field beamspace are similar to those of the far-field beamspace in the left of Fig.~\ref{BeamSpace}. Specifically, the 3~dB beamwidth of the high mainlobe along the angle is approximately two divided by the number of antennas, which indicates that the ability of beam focusing in the angle dimension increases linearly with the number of antennas.

\textit{Distance Cross-Section:} In  the right of Fig.~\ref{NearfieldBeamFocusing}, we illustrate the distance cross-section of the near-field beamspace across the focused location. Since the distance cross-section passes through the focused location, the  angle of the FrFT basis aligns with the linear coefficient of the channel steering vector. In this context, the FrFT of the channel steering vector would be the summation of quadratic phase signals. By approximating the summation as the integral and employing the Fresnel function~\cite{Tcom22CMH}, we can determine that the 3~dB beamwidth of the high mainlobe along the distance is approximately seven divided by the square of the number of antennas. Interested readers can refer to \cite{JSTSP24CKJ} for more details.  This outcome  indicates that the ability of beam focusing in the distance dimension increases quadratically with the number of antennas.

\textit{Overall High Mainlobe:} Inspired by the analogous Taylor series expansions of the Gaussian function and the beamforming gain within the high mainlobe, we approximate the high mainlobe using a Gaussian function, which is referred to as the Gaussian approximation~\cite{TWC24CKJ2}. From \cite{TWC24CKJ2}, this approximation is highly accurate, with an average deviation of only 0.5\%. This approach provides a clear and comprehensive understanding of the high mainlobe. Based on the Gaussian approximation, the beam gain contours of the high mainlobe can be expressed as elliptic functions. Then, the analyzed properties of the angle cross-section and the distance cross-section can be extended to the overall high mainlobe.

\begin{table*}[!ht]

	\centering
	\caption{Comparisons of  Different Methods  that Explore the Channel Properties of XL-MIMO Systems.}\label{tabexample}
	\begin{tabular}{|c|c|c|c|c|c|c|c|c|}
		\hline
		\multirow{2}{*}{\makecell[c]{Methods\\}} & \multicolumn{2}{c|}{Channels} & \multicolumn{3}{c|}{Beam Pattern} & \multicolumn{3}{c|}{Properties}  \\ \cline{2-9}
		~ & Far Field & Near Field & High Mainlobe &Low Mainlobe &  Analysis & PR & TI & FC \\ \hline
		Far-Field Beamspace\cite{BSMIMO} & \raisebox{0.6ex}{\scalebox{0.7}{$\sqrt{}$}} &\scalebox{0.85}[1]{$\times$} & \scalebox{0.85}[1]{$\times$} & \scalebox{0.85}[1]{$\times$} & \scalebox{0.85}[1]{$\times$} & \raisebox{0.6ex}{\scalebox{0.7}{$\sqrt{}$}} & \scalebox{0.85}[1]{$\times$} & \raisebox{0.6ex}{\scalebox{0.7}{$\sqrt{}$}}  \\ \hline
		Near-field Beam Focusing~\cite{TWC22ZHY} & \raisebox{0.6ex}{\scalebox{0.7}{$\sqrt{}$}} & \raisebox{0.6ex}{\scalebox{0.7}{$\sqrt{}$}} & \raisebox{0.6ex}{\scalebox{0.7}{$\sqrt{}$}} & \scalebox{0.85}[1]{$\times$} & \scalebox{0.85}[1]{$\times$} & \scalebox{0.85}[1]{$\times$} & \scalebox{0.85}[1]{$\times$} & \scalebox{0.85}[1]{$\times$}  \\ \hline
		Polar domain~\cite{Tcom22CMH} & \raisebox{0.6ex}{\scalebox{0.7}{$\sqrt{}$}} & \raisebox{0.6ex}{\scalebox{0.7}{$\sqrt{}$}} & \raisebox{0.6ex}{\scalebox{0.7}{$\sqrt{}$}} & \raisebox{0.6ex}{\scalebox{0.7}{$\sqrt{}$}} & $\bigtriangleup$ & \scalebox{0.85}[1]{$\times$} & \scalebox{0.85}[1]{$\times$} & \scalebox{0.85}[1]{$\times$}  \\ \hline
		$k-b$ domain~\cite{TWC23SX} & \raisebox{0.6ex}{\scalebox{0.7}{$\sqrt{}$}} & \raisebox{0.6ex}{\scalebox{0.7}{$\sqrt{}$}} & \scalebox{0.85}[1]{$\times$} & \raisebox{0.6ex}{\scalebox{0.7}{$\sqrt{}$}}& $\bigtriangleup$ & \raisebox{0.6ex}{\scalebox{0.7}{$\sqrt{}$}} & \scalebox{0.85}[1]{$\times$} & \scalebox{0.85}[1]{$\times$}  \\ \hline
		Wavenumber domain~\cite{CM24CYB}  & \raisebox{0.6ex}{\scalebox{0.7}{$\sqrt{}$}} & \raisebox{0.6ex}{\scalebox{0.7}{$\sqrt{}$}} & \scalebox{0.85}[1]{$\times$} & \scalebox{0.85}[1]{$\times$} & \scalebox{0.85}[1]{$\times$} & \scalebox{0.85}[1]{$\times$} & \scalebox{0.85}[1]{$\times$} &  \raisebox{0.6ex}{\scalebox{0.7}{$\sqrt{}$}} \\ \hline
		Near-Field Beamspace & \raisebox{0.6ex}{\scalebox{0.7}{$\sqrt{}$}} & \raisebox{0.6ex}{\scalebox{0.7}{$\sqrt{}$}} & \raisebox{0.6ex}{\scalebox{0.7}{$\sqrt{}$}} & \raisebox{0.6ex}{\scalebox{0.7}{$\sqrt{}$}} & \raisebox{0.6ex}{\scalebox{0.7}{$\sqrt{}$}} & \raisebox{0.6ex}{\scalebox{0.7}{$\sqrt{}$}} & \raisebox{0.6ex}{\scalebox{0.7}{$\sqrt{}$}} & \raisebox{0.6ex}{\scalebox{0.7}{$\sqrt{}$}} \\ \hline
		\multicolumn{9}{|l|}{\footnotesize \textbf{Note:} The  symbol  \raisebox{0.6ex}{\scalebox{0.7}{$\sqrt{}$}} indicates that the corresponding criterion is addressed in the method; ~The  symbol \scalebox{0.85}[1]{$\times$} indicates that the corresponding criterion}\\ 
		\multicolumn{9}{|l|}{\footnotesize~~~~~~~~is not addressed in the  method; The symbol $\bigtriangleup$  indicates that the corresponding criterion is partially addressed in the method.}\\ 
		\hline
	\end{tabular}
	\vspace{-0.4cm}
\end{table*}

\subsubsection{\textbf{Low Mainlobe}}\label{LowMainlobeA} Now we analyze the characteristics of the low mainlobe in the near-field beamspace. From the near-field beamspace in the right of Fig.~\ref{BeamSpace}, the beamwidth of the low mainlobe along the angle increases with the deviation between the surrogate distances of the focused location and an arbitrary location, which is termed as the surrogate distance deviation for simplicity; and the beamforming gain of the low mainlobe decreases with the surrogate distance deviation. Understanding the variations of the beamwidth and beamforming gain with the  surrogate distance deviation is  the key to exploring the characteristics of the low mainlobe. 

\textit{Spatial Region and Intensity Distribution of the Low Mainlobe:} In the low mainlobe, there is a misalignment between the FrFT basis and the near-field channel steering vector. As a result, given a fixed location, the beamforming gain along the angle can be seen as the Fourier transform of quadratic phase signals~\cite{TWC23SX}. However, deriving closed-form expressions for this problem is challenging due to the quadratic phase terms. Numerical methods may provide well-approximated results  for this problem but cannot provide enough insights for the understanding of the low mainlobes. Note that the classic principle of stationary phase (PSP) provides a concise and insightful analysis for the spectrum of the quadratic phase signals~\cite{TWC23SX,TWC24CKJ2}. Based on the PSP, the beamwidth of the low mainlobe along the angle increases linearly with the increase of the surrogate distance deviation, while the square of the average beamforming gain of the low mainlobe increases inversely with  the increase of  the surrogate distance deviation. This succinct relation provides an effective and intuitive approach to characterize  the low mainlobe of the near-field beamspace, which can facilitate several applications, as will be stated in Section~\ref{NFCSIAcq}.  In Fig.~\ref{LowMainlobe}, we illustrate the spatial region and intensity distribution of the low mainlobe in the near-field beamspace. From the figure, the calculated average beamforming gain and beamwidth show a striking similarity to their approximations based on the PSP. From Fig.~\ref{LowMainlobe}, the approximated beamforming gain deviates by less than 2\% from the calculated value, and the approximated beamwidth deviates by less than the inverse of the number of antennas from the calculated value. This similarity underscores the effectiveness of the PSP-based approximation in capturing the characteristics of the low mainlobe.

% As depicted in Fig.~\ref{LowMainlobe}b), the calculated average beam gain and beamwidth show a striking similarity to their approximations based on the PSP. This similarity underscores the effectiveness of the PSP-based approximation in capturing the characteristics of the low mainlobe, as discussed earlier.

%\textit{Angle Cross-Section of the Low Mainlobe:} Fig.~\ref{LowMainlobe}c) illustrates the angle cross-section of the low mainlobe, where $b = 1.3\times 10^{-4}$. We compare the real beam pattern with the approximate beam pattern based on the PSP in terms of the beam gain and the beamwidth. From the figure, the real beam pattern and the approximated one are similar, which indicates that the PSP can well approximate the low mainlobe of the near-field beamspace.

\subsection{Properties of the Near-Field Beamspace}\label{PNFB}
The near-field beamspace is obtained by applying the FrFT to the near-field channel steering vector, which allows leveraging FrFT properties to enhance the analysis and computation of the near-field beamspace.

\textit{Parseval’s Relation (PR) in the Near-Field Beamspace:}  A significant property of the FrFT is its adherence to the Parseval’s theorem~\cite{TSP94ALB}. Based on this property, we can obtain that the accumulated energy of the near-field beamspace is the same for different surrogate  distances. In far-field scenarios, the beamforming gain deteriorates when the beam is directed towards multiple distinct directions compared to pointing to a single direction. However, in the near field, due to this property, the BS can provide beamforming gains for locations in low mainlobes without degrading the beamforming gain of the focused location.

\textit{Translation Invariance (TI) in the Near-Field Beamspace:} The previous analysis of the near-field beamspace is based on a specific channel steering vector, which may limit its general applicability. Thankfully, as demonstrated in \cite{JSTSP24CKJ}, the FrFT of the near-field channel steering vector exhibits the property of translation invariance, which  means that a change in the channel steering vector results in a corresponding translation of the near-field beamspace.  In this condition, the focused location can change without affecting the characteristics of the mainlobe and sidelobe in the near-field beamspace.  Consequently, the properties of the mainlobe and sidelobes analyzed for one channel steering vector can be directly extended to any other channel steering vector.

\textit{Fast Calculation (FC) of the Near-Field Beamspace:} In XL-MIMO systems, the antenna array typically consists of a large number of elements, resulting in high-dimensional channel steering vectors. Moreover, the FrFT is a two-dimensional transform involving both angle and distance, further complicating the near-field beamspace analysis. To alleviate this complexity, the fast FrFT algorithm, which achieves the same order of computational complexity as the fast Fourier transform, can be employed~\cite{swarztrauber1991fractional}. This approach  significantly enhances computational efficiency and supports timely processing, making it highly suitable for practical near-field beamspace analysis.

\subsection{Comparisons with the Existing Works}
Existing works have explored various methods to analyze the channel properties of XL-MIMO systems,  including the far-field beamspace \cite{BSMIMO}, near-field beam focusing \cite{TWC22ZHY}, polar domain \cite{Tcom22CMH}, $k-b$ domain \cite{TWC23SX}, and wavenumber domain \cite{CM24CYB}. Table \ref{tabexample} compares the near-field beamspace with these methods based on eight criteria: applicability to far-field channels, applicability to near-field channels, focus on the high mainlobe, focus on the low mainlobe, inclusion of analysis for both high and low mainlobes, adherence to PR, consideration of TI, and capability for FC. The table shows that the proposed near-field beamspace offers a more comprehensive approach to analyzing near-field channels compared to the existing methods, highlighting its effectiveness in addressing the complex characteristics of near-field channels in XL-MIMO systems.

\section{Applications of Near-Field Beamspace}\label{NFCSIAcq}
The near-field beamspace analysis has revealed several properties that are not addressed by the existing works. By exploiting these properties, several promising applications can be developed. 
%In the following, we will provide a few examples, with a multitude of other scenarios warranting further investigation.

\subsection{Near-Field Beam Training}\label{NFBTT}
As an effective way to accomplish the beam alignment, beam training  explores the communication channels by testing predefined codewords. In the context of near field, a straightforward method to perform beam training  is the near-field beam sweeping~\cite{WCL22ZYP}. This approach first samples the space in angle and distance, and then leverages the  near-field beam focusing to evaluate each sampled position. However, its training overhead depends on the samples in both angle and distance and therefore is extremely high. For example, with 512 angle samples and 11 distance samples~\cite{TWC23CKJ}, the training overhead will be $512\times11 = 5632$.  In fact, the high training overhead of the near-field beam sweeping is attributed to the exclusive  emphasis on high mainlobes and neglect of the low ones.  From Section \ref{LowMainlobeA}, the low mainlobe can also provide considerable beamforming gain and  has much wider beam coverage than the high one, which indicates much fewer codewords are needed to cover the whole near-field beamspace. In \cite{TWC24CKJ2}, only 15 codewords are needed to cover the space for beam training, which is much less than the  beam sweeping, e.g. 15 v.s. 5632. Therefore, the near-field training overhead can be substantially reduced by  exploiting the low mainlobes and  carefully designing the codebook~\cite{TWC23SX}.

\subsection{Near-Field Beam Refinement}\label{NFBR}
One fundamental problem with beam training is the limited resolution due to the quantization. A straightforward way to improve the estimation accuracy is to reduce the quantization intervals. However, the smaller quantization intervals will inevitably lead to a larger codebook size, and a higher training overhead is needed to test the codewords in the enlarged codebook. In this condition, the low-complexity beam refinement can be considered~\cite{TWC23CKJ}. From Fig.~\ref{NearfieldBeamFocusing}, the high mainlobe of the beam pattern can be highly approximated as a two-dimensional Gaussian function. By exploiting this property, the parameter estimation of the near-field channels can be converted to the parameter estimation of the two-dimensional Gaussian function. Based on the well-explored parameter estimation of  Gaussian functions, closed-form estimates of channel angle and distance can be derived~\cite{TWC24CKJ2}. Then, the estimated parameters can be exploited to improve positioning performance and achieve accurate beam alignment.

%Fig.~\ref{PotentialApplications}b) illustrates the performance of the beam refinement based on the Gaussian approximation. From the figure, the beam refinement has great advantages over beam training in both positioning error and beamforming gains.

\subsection{Near-Field Beam Tracking}\label{NFBT}
To further reduce beam alignment overhead, a near-field beam tracking method has been developed~\cite{TWC23CKJ}. This approach leverages the correlation of channel state information across different time slots to estimate real-time user positions. During  beam tracking, when a user moves beyond the mainlobe coverage, the BS adjusts the beamforming vectors to maintain high beamforming gain. A crucial step in this process is to determine the extent of the high mainlobe coverage. By utilizing the Gaussian approximation, the detailed coverage of the high mainlobe in all directions can be conveniently assessed. Based on this assessment, the beams can be adaptively adjusted to achieve effective near-field beam tracking. These enhancements to beam tracking are essential for optimizing performance in beam alignment.

\subsection{Near-Field Sensing and  Localization}\label{NFS}
Besides the conventional communication ability, future wireless networks can also utilize ubiquitous communication signals to sense surrounding targets. In the context of XL-MIMO, changes in channel models  have shifted the focus to  the near-field sensing and localization~\cite{TSP21GA}. Previous works have demonstrated that, near-field beam focusing can localize targets based solely on spatial information by exploiting the spherical wavefront. In addition to the focused location in near-field beam focusing, attention should also be given to the wide-coverage low mainlobe.  On one hand, the low mainlobe may have higher received power than the high mainlobe if the targets in the low mainlobe are closer to the BS than the focused location.  On the other hand, the radiating of the low mainlobe will not deteriorate the performance of the high mainlobe due to the Parseval’s relation in the near-field beamspace.  Consequently, utilizing the high and low mainlobes in near-field beamspace is essential for enhancing the sensing capabilities of XL-MIMO systems.

\subsection{Near-Field Channel Estimation}\label{NFCE}
XL-MIMO systems typically employ a large number of antennas,  which results in high-dimensional channel matrices that present significant challenges for channel estimation. To reduce the estimation complexity, the channel sparsity can be exploited.  Since the beam energy can be concentrated via the FrFT, the near-field channels exhibit sparsity in the near-field beamspace. Due to the translation invariance property, the sparsity is uniform throughout the beamspace, which is significantly different from the polar-domain sparsity~\cite{Tcom22CMH} that has uniform sparsity in angle but nonuniform sparsity in distance. Leveraging this uniform sparsity, efficient sparse channel estimation algorithms can be developed~\cite{JSTSP24CKJ}.

\subsection{Near-Field Efficient Beamforming Strategies}\label{NFFB}
Based on the analysis on the spatial region and intensity distribution of the mainlobes in the near-field beamspace, more efficient beamforming strategies can be devised. For instance, in multiuser communications, low-complexity user grouping can be efficiently implemented by grouping users with low channel coherence through near-field beamspace analysis. Additionally, in beam allocation, when beam conflicts among users arise, beam scheduling can be adjusted to balance beamforming gain and multiuser interference based on the beam pattern analysis in the near-field beamspace.

\section{Conclusion and Future Directions}\label{Conclu}
Different from traditional communication systems that operate in the far-field regime, the evolution of XL-MIMO systems has introduced the near-field communications. In this article, we have investigated the near-field communications for XL-MIMO from the perspective of beamspace. We have pointed out that the antenna space and the near-field beamspace  should be connected by the FrFT instead of the commonly-used Fourier transform. Then, we have divided the near-field beamspace into the high mainlobe, the low mainlobe and the sidelobe. Based on the comprehensive analysis on the near-field beamspace, we have presented several applications of the near-field beamspace that are not addressed by the existing works.

Note that the research on the near-field beamspace is still in its initial stage compared to the well-explored far-field one. In the following, we highlight some important challenges and potential directions for future study. 
	
\textit{Near-Field Beamspace for Wideband Communications:} In millimeter wave or terahertz systems, which typically employ large bandwidth, the near-field beamspace will encounter the beam squint or beam split effects. Different carrier frequencies result in different properties for both the high and low mainlobes across different subcarriers, where the different properties can be distinct for wideband communications. However, the connections among the near-field beamspace for different subcarriers remain unknown, which is worthy of further exploration.

\textit{Near-Field Beamspace for Double-Side XL-MIMO:} In the far field, the beamspace of double-side massive MIMO can be decomposed into the combination of the beamspace for both the transmitter and the receiver. However, this relation does not hold for the double-side XL-MIMO because of the spherical wavefront in the near field. Therefore, extending the near-field beamspace for single-side XL-MIMO to that for double-side XL-MIMO is not straightforward, which calls for specialized studies on the properties of the near-field beamspace of double-side XL-MIMO.

\textit{Near-Field Beamspace for Reconfigurable Intelligent Surfaces (RIS)-Aided Wireless Communications:}  RIS has recently gained significant attention for its ability to modify the electromagnetic environment in wireless communications. As the size of RIS increases, the near-field region also expands, covering a substantial portion of the service area. In this context, analyzing the near-field beamspace becomes crucial for efficient beamforming in RIS-assisted systems. Unlike traditional communication channels, RIS channels require consideration of both incident and reflected signals, necessitating a near-field beamspace specifically designed for RIS applications.

\textit{Artificial Intelligence (AI)-Empowered Approaches:} The exploration of the near-field beamspace together with the corresponding signal processing techniques can improve the beamforming efficiency and flexibility. However, the complicated features of the near-field beamspace pose significant challenges to classic methods. Fortunately, leveraging the powerful learning and prediction capabilities of AI especially for nonlinear problems may offer promising approaches to further explore the near-field communications.

\bibliographystyle{IEEEtran}
\bibliography{IEEEabrv,IEEEexample}

\section*{Biographies}

\begin{IEEEbiographynophoto}{Kangjian Chen}
 is  currently pursuing the Ph.D. degree in signal processing at Southeast University, Nanjing, China. 
\end{IEEEbiographynophoto}

\begin{IEEEbiographynophoto}{Chenhao Qi [SM]}
 is currently a Professor with the School of Information Science and Engineering, Southeast University, Nanjing, China.
\end{IEEEbiographynophoto}

\begin{IEEEbiographynophoto}{Jingjia Huang}
	is  currently pursuing the Ph.D. degree in signal processing at Southeast University, Nanjing, China. 
\end{IEEEbiographynophoto}
\begin{IEEEbiographynophoto}{Octavia A. Dobre [F]}
is currently a Professor and Canada Research Chair Tier 1 with Memorial University, Canada.
\end{IEEEbiographynophoto}
\begin{IEEEbiographynophoto}{Geoffrey Ye Li [F]}
 is currently a Chair Professor at Imperial College London, UK.
\end{IEEEbiographynophoto}
% that's all folks
\end{document}